\documentstyle[12pt]{article}

\begin{document}

\newcommand{\non}{\nonumber}
\newcommand{\s}{\\ \vspace*{-2mm}}
\newcommand{\nn}{\noindent}
\newcommand{\nt}{\not \hspace*{-1mm} }
\newcommand{\be}{\begin{eqnarray}}
\newcommand{\en}{\end{eqnarray}}

\renewcommand{\thefootnote}{\fnsymbol{footnote} }

\nn \hspace*{12cm} UdeM-LPN-TH-93-156 \\
    \hspace*{12cm} NYU--TH--93/05/02 \\
\hspace*{12cm} July 1993

\vspace*{2cm}

\centerline{\large{\bf On the asymptotic ${\cal O}(\alpha \alpha_S)$ behavior
of the electroweak gauge bosons}}

\vspace*{0.4cm}

\centerline{\large{\bf vacuum polarization functions for arbitrary quark
masses.}}

\vspace*{1.5cm}

\centerline{\sc A.~Djouadi$^1\footnote{NSERC Fellow.}$ and P.~Gambino$^2$.}

\vspace*{1cm}

\centerline{$^1$ Laboratoire de Physique Nucl\'eaire, Universit\'e de
Montr\'eal,  Case 6128 Suc.~A,}
\centerline{H3C 3J7 Montr\'eal PQ, Canada.}

\vspace*{0.4cm}

\centerline{$^2$ Department of Physics, New York University, 4 Washington
Place,}
\centerline{ New York, NY 10003, USA.}

\vspace*{2cm}

\begin{center}
\parbox{14cm}
{\begin{center} ABSTRACT \end{center}
\vspace*{0.2cm}

\nn We derive the QCD corrections to the electroweak gauge bosons vacuum
polarization functions at high and zero--momentum transfer in the case of
arbitrary internal quark masses. We then discuss in this general case
$(i)$ the connection between the ${\cal O}(\alpha \alpha_S)$ calculations of
the vector bosons self--energies using dimensional regularization and the one
performed via a dispersive approach and $(ii)$ the QCD corrections to the
$\rho$ parameter for a heavy quark isodoublet.}

\end{center}

\newpage

\renewcommand{\thefootnote}{\arabic{footnote} }
\setcounter{footnote}{0}

\subsection*{1.~Introduction}

The Minimal Standard Model of the electromagnetic, weak and strong interactions
[with three generations of fermions and one Higgs doublet] has achieved a
tremendous success in describing all experimental data within the range of
energies available today. In particular, it has been tested to the level of its
quantum corrections in the high--precision LEP and SLC experiments \cite{lep}.
To allow for such precision tests, a considerable amount of theoretical work
has been done in the last few years to calculate the relevant radiative
corrections \cite{zp}. \s

\nn Among these radiative corrections, a special attention has been devoted to
the effects of the top--bottom weak isodoublet in the vacuum polarization
functions of the electroweak gauge bosons. Virtual top quarks appear in these
self--energies and lead to contributions that are quadratically proportional to
the large mass of this still undiscovered particle \cite{mt}. These leading
quadratic corrections can be parametrized by a deviation from unity of the
$\rho$ parameter \cite{vel} which measures the relative strength of the neutral
to charged currents at zero--momentum transfer. \s

\nn In fact, due to the large size of these contributions and in view of the
remarkably high accuracy of the experimental data, some corrections were
required beyond the one--loop approximation. The first one involves virtual
Higgs boson exchange which leads to potentially large ${\cal O}(\alpha^2
m_t^4/M_W^4)$ contributions \cite{higgs,barb}. Motivated by the large value of
the QCD coupling constant, the second type of corrections are the two--loop
${\cal O} (\alpha \alpha_S)$ contributions which have been calculated in
\cite{moi} using dimensional regularization and in \cite{kn} via a dispersive
approach; the connection between these two calculations [which is needed in
order to implement non-perturbative effects in various renormalization schemes]
has been discussed recently in \cite{ks,ksf}. \s

\nn Once the top quark is found, a comparison of the experimental data with the
accurate theoretical predictions will allow to search for virtual effects of
New Physics beyond the Minimal Standard Model. An immediate candidate would be
a fourth generation of fermions, the existence of which is still allowed by
present data if the associated neutrino is heavy enough \cite{lep}. The latter
would contribute to the vector bosons self--energies in a way similar to the
t--b isodoublet. For instance, if the mass splitting between the two quarks is
large, the main contribution will be cast into a correction to the $\rho$
parameter. \s

\nn However, the previously mentioned two--loop results cannot be extrapolated
to this case: due to the technical difficulties that one encounters when
calculating at this level of perturbation theory, these results involve the
simplifying approximation that the bottom quark is massless. [Note that this is
not the case of the Higgs correction to the $\rho$ parameter derived in
\cite{higgs} where the full dependence on $m_b$ has been retained, but the
Higgs boson mass has been neglected, an approximation which turned out to fail
badly \cite{barb} for realistic values of the quark masses]. While this
approximation is certainly good enough for the top--bottom isoudoublet, it is
very unlikely that this will be the case for extra quark isodoublets. \s

\nn In this paper, we calculate using dimensional regularization the two--loop
QCD corrections to the electroweak gauge bosons vacuum polarization functions
at high--momentum and at zero--momentum transfer in the case of arbitrary quark
masses. This allows to derive the ${\cal O}(\alpha \alpha_S)$ corrections to
the $\rho$ parameter, to which heavy quark isodoublets will mainly contribute
if there is a sizeable splitting between the two quark masses. The full QCD
corrections, including the subleading terms, can be obtained via a [numerical]
dispersive integral of the imaginary parts of the vector boson self--energies
which are known in the case of arbitrary quark masses \cite{ima}. Using the
high--momentum transfer behavior that we derive here, one can obtain the vacuum
polarization functions in dimensional regularization; this allows to implement
non--perturbative effects in any renormalization scheme. \s

\nn The paper is organized as follows. In the next section we summarize the
one--loop results which will be relevant to our discussion. In section 3, we
discuss the high--energy behavior of the vacuum polarization functions and
generalize to the case of arbitrary quark masses and at the two--loop level,
recent analyses of the connection between calculations using dimensional
regularization and the one using dispersive integration \cite{ks}. In section
4, we discuss the behavior of the vacuum polarization functions at
low--momentum transfer and derive the QCD corrections to the $\rho$ parameter
in the case of a heavy quark isodoublet. Section 5 contains our conclusions.

\subsection*{2.~One--loop results}

\renewcommand{\theequation}{2.\arabic{equation}}
\setcounter{equation}{0}

\nn To set the notation and for the sake of completeness, we summarize in this
section all the one--loop results which will be relevant to our discussion. \s

\nn The contribution of a fermionic loop to the vacuum polarization tensor of a
vector boson $i$, or to the mixing amplitude of two bosons $i$ and $j$, denoted
$\Pi_{\mu\nu}^{ij}$, is defined as
\be
\Pi_{\mu\nu}^{ij}(q^2)= -i \int d^4 x e^{i q\cdot x}<0|{\rm T}^* \left[J_\mu^{
i}(x) J_\nu^{j \dagger }(0) \right]|0>
\label{def} \en
where T$^*$ is the covariant time ordering product and $q$ the four--momentum
transfer; $J_\mu^{i}, J_\nu^{j}$ are fermionic currents coupled to the vector
bosons $i,j$ and constructed with spinor fields whose corresponding masses are
$m_a, m_b$. The vacuum polarization tensor can be decomposed into a transverse
and a longitudinal part,
\be
\Pi_{\mu\nu}^{ij}(q^2)=\left(g_{\mu \nu}-\frac{q_\mu q_\nu}{q^2}
\right) \Pi_T^{ij}(q^2)\ + \ \frac{q_\mu q_\nu}{q^2}\ \Pi_L^{ij}(q^2)
\en
and the two components can be directly extracted by contracting $\Pi_{\mu \nu
}^{ij} (q^2)$ by the Lorentz invariants $g_{\mu \nu}-q_\mu q_\nu/q^2$ and
$q_\mu q_\nu/q^2$; they can be written as [with $s \equiv q^2$]
\be
\Pi_{T,L}^{ij}(s) \ = \ \frac{\alpha}{\pi} \ \left[ \ v^iv^j \ \Pi_{T,L}^V (s)
\ + \ a^ia^j \ \Pi_{T,L}^A (s)\
\right]
\en
with $v^i$ and $a^i$ the vector and axial--vector couplings of the gauge boson
$i$ to the fermions expressed in units of the proton charge $e=\sqrt{4 \pi
\alpha}$. Note that, for amplitudes involving very light external fermions,
only
the transverse part of the self--energies contribute significantly. \s

\nn The vacuum polarization function can be calculated directly by evaluating,
for instance at the one--loop level, the diagram of Fig.~1a using dimensional
regularization. Alternatively, the polarization function can be expressed as a
dispersive integral of its imaginary part [which is related to the decay widths
of the vector bosons into fermions],
\be
\Pi(s, \Lambda) = \frac{1}{\pi} \ \int_{(m_a+m_b)^2}^{\Lambda^2} {\rm d}
s' \ \frac{ {\cal I}m \Pi(s) }{s-s'-i\epsilon }
\label{4.1} \en
where the cut--off scale $\Lambda \gg\sqrt{s},m_a,m_b$ regulates the
ultraviolet divergence. However, in order to be  meaningful, the previous
dispersion relation needs to be subtracted.  For a conserved current, as in the
case of QED, the presence of spurious quadratic divergence is removed by a
subtraction of the integral at $s=0$, restoring in this way the Ward identity
which requires the photon vacuum polarization function to be transverse and
proportional to $s$. The situation becomes more complicated for a
non--conserved current, for which some logarithmic divergencies are expected to
be proportional to the squared fermion masses, and should not be removed in the
subtraction procedure. \s

\nn Recently, Kniehl and Sirlin proposed a subtraction prescription for the
dispersion relation eq.~(2.4) which manifestly preserves the generic Ward
identities constraining the vacuum polarization functions \cite{ks}. Let us
recall the basic elements of this procedure and write the Ward identity
\be
q_\nu\Delta(s) \equiv \int d^4x e^{iq\cdot x}<0|{\rm T}^*\left[ \partial^\mu
J_\mu (x)J_\nu^{\dagger}(0)\right]|0> =q^\mu \Pi_{\mu \nu}(s)
\en
with a notation as in eq.~(2.1) but where we dropped the indices $i,j$
for simplicity. Contracting the previous equation by $q^\nu$, one obtains
$\Delta (s) = \Pi_L(s)$; then using the linear combination $s\Phi(s)= \Pi_L(s)-
\Pi_T(s)$, one can rewrite the transverse part in terms of $\Delta(s)$ and
$\Phi(s)$,
\be \Pi_T(s)= -s \Phi(s)+ \Delta(s) \hspace*{0.5cm} , \hspace*{0.8cm}
\Delta (s) = \Pi_L(s) \label{wi}\en
Now $\Phi (s)$ and $\Delta(s)$ are only logarityhmically divergent for infinite
energies, unlike $\Pi_T(s)$  which is quadratically divergent as discussed
above. For $\Phi (s)$ this is obvious for dimensional reasons [two powers of
the external momentum have been extracted], and for $\Delta(s)$ this is because
the soft breaking of the current $J_\mu$ by mass terms implies that its
derivative involves at most operators of canonical dimension three, and one
power of the momentum is extracted. \s

\nn Following the authors of Ref.~\cite{ks}, we therefore write the once
subtracted dispersion relation for the transverse part of the vacuum
polarization function expressing  $\Pi_T (s)$ through eq.~(\ref{wi}), and then
writing unsubtracted dispersion relations for  $ \Pi_L (s)$ and $\Phi (s)$,
\be
\Pi_T(s)= {1\over\pi} \int_{(m_1+m_2)^2}^{\Lambda^2} ds' \left[ {{\cal I}
m\Pi_T(s') \over {s' -s-i\epsilon}}+ {\cal I}m\Phi(s') \right] +  {1\over2\pi
i} \oint_{|s'|=\Lambda^2}{ds'\over s'} \left[ \Pi_L(s')-s\Phi(s') \right]
\label{dr}
\en
where the quadratic divergence is now cancelled by the contribution of
${\cal I}m\Phi(s')$,
and the logarithmic divergence is  regulated by the cut--off $\Lambda$.
The integral over the large circle, where the $s$ in the denominator has been
neglected compared to $s'$, comes from the fact that $\Pi_L(s)$ and
$\Phi(s)$ do not vanish asymptotically but behave like constants modulo
logarithms. Note that for conserved currents $\partial_\mu  J^\mu(x)=0$,
$\Pi_L(s)$ vanishes reducing then the Ward identity to the QED expression
$\Pi_T(s)=-s\Phi(s)$. \s

\nn One can now express the result of the dispersion integral eq.~(2.7) for
arbitrary quark masses and for vector and axial--vector currents, as
\be
\Pi_{T}^{V,A} (s) = sX_1 +(m_a \mp m_b)^2 Y_1 + F_1^{V,A}(s)\ +\frac{1}{2\pi i}
\oint_{|s'|=\Lambda^2}\frac{{\rm d}s'}{s'} \left[ \Pi_L^{V,A} (s')
-s\Phi^{V,A} (s') \right] \label{disp}
\en
where the function $F^{V,A}_1(s)$  behaves like a constant as $s\to
\infty$. $X_1$ and $Y_1$ are divergent constants involving logarithms of the
cut--off scale $\Lambda$; their explicit form is given by
\be
X_1 = \log\frac{\Lambda^2}{m_am_b} -\frac{5}{3} \ , \hspace*{0.8cm}
Y_1 = -\frac{3}{2} \log \frac{\Lambda^2}{m_am_b} + 3 \
\en

\nn Analogously, the real part of $\Pi_T^{V,A}(s)$ calculated in $n=4-2\epsilon
$ dimensions can be written as
\be
\Pi_{T}^{V,A}(s)= s \tilde{X}_1 +(m_a \mp m_b)^2 \tilde{Y}_1 + F_1^{V,A} (s)
\label{dimreg}
\en
where now, the divergent constants $\tilde{X}_1$ and $\tilde{Y}_1$ involve
poles in $\epsilon=2-n/2$ and logarithms of the 't~Hooft mass scale $\mu$;
their expressions at the one--loop level are given by
\be
\tilde{ X}_1 = \frac{1}{\epsilon} - \log\frac{m_am_b}{\mu^2} \ ,
\hspace*{0.8cm}
\tilde{ Y}_1 = - \frac{3}{2\epsilon} + \frac{3}{2} \log \frac{m_am_b}{\mu^2} \
\en
\nn The finite function $F_1^{V,A}(s)$ in eq.~(2.8) is, of course, the same as
in eq.~(2.10). Since the result of eq.~(\ref{dr}) should be consistent with the
one obtained using dimensional regularization, the integral over the large
circle must be absorbed by a redefinition of the divergent constants $X_1$ and
$Y_1$. In particular, if $\Pi_L(s)$ and $\Phi(s)$ are evaluated in dimensional
regularization, the effect of the second integral in eq.~(\ref{dr}) should be
simply to replace $X_1,~Y_1$ by $\tilde{X}_1, ~\tilde{Y}_1$. This is indeed
true and one has
\be
\frac{1}{2\pi i}\oint_{|s'|=\Lambda^2}ds' \frac{\Phi^{V,A}(s')}{s'} &=&
X_1 - \tilde{X}_1 \non \\
\frac{1}{2\pi i}\oint_{|s'|=\Lambda^2}ds' \frac{\Pi^{V,A}_{L}(s')}{s'}&=&
(m_a \mp m_b)^2 ( \tilde{Y}_1 -Y_1 ) \label{largecircle}
\en
\nn This point is crucial to ensure that the cancellation of divergences
occurring in physical observables like the $\rho$ parameter yields a perfect
independence of these observables from the calculational approach used to
obtain the vacuum polarization functions. \\
Before we generalize this analysis for arbitrary quark masses at the
two--loop level, let us briefly discuss the contribution of a heavy quark
isodoublet to the $\rho$ parameter. \s

\nn The fermionic contribution to the $\rho$ parameter can be expressed in
terms of the difference between the transverse components of the $Z$ and $W$
bosons self--energies at zero--momentum transfer, $q^2=0$
\be
\Delta \rho \ = \ \frac{1}{\rho} -1 \ = \ \frac{\Pi^{ZZ}_T(0)}{M_Z^2} -
\frac{\Pi^{WW}_T(0)}{M_W^2}
\en
Besides the light fermions contribution to the photon vacuum polarization
function which can be mapped into the running of the fine structure constant
$\alpha$, $\Delta \rho$ represents the leading part of the radiative correction
to electroweak observables such as the $W$ boson mass and the effective weak
mixing angle $\sin^2\theta_W$ as measured for instance on the $Z$ resonance
\cite{zp}. \s

\nn The contribution of a quark pair with different masses, $m_a \neq m_b \neq
0$, to $\Pi^{V,A}_{T,L}(0)$ in the one--loop approximation, Fig.~1a, is
\be
\Pi^{V,A}_{T,L} (0) &=& - \frac{3}{4} \left[ (m_a^2+m_b^2) \left(\frac{2}
{\epsilon} + 1 - \rho_a - \rho_b \right) + \frac{m_a^4+ m_b^4}{m_a^2-m_b^2}
\log \frac{m_b^2}{m_a^2} \right] \non \\
& & \pm \ 3 m_a m_b \left[ \frac{1}{\epsilon} +1 - \frac{1}{2}(\rho_a+\rho_b)
-\frac{1}{2} \frac{m_a^2+m_b^2}{m_a^2-m_b^2} \log \frac{m_a^2}{m_b^2} \right]
\en
with $\rho_{a,b}= \log m_{a,b}^2/\mu^2$ where $\mu$ is the 't Hooft mass scale
introduced to make the coupling constants dimensionless in $n=4-2\epsilon$
dimensions [note that we have introduced an extra term $(e^\gamma/4\pi)^
\epsilon$, where $\gamma$ is the Euler constant, to prevent uninteresting
combinations of $\log 4\pi, \gamma, \cdots$ in the previous result]. One can
easily verify that for a vector current with equal fermion masses $m_a=m_b$, as
it is the case for the photon or the $\gamma$--$Z$ mixing amplitude, the vacuum
polarization function vanishes at $q^2=0$. \s

\nn In the case of a heavy quark isodoublet [with Standard Model couplings]
$\Delta \rho$ can alternatively be written in terms of the vector and
axial--vector components of the vacuum polarization function eq.~(2.3);
using the fact that $\Pi^V(0,m_a,m_a)$ vanishes and that in this
limit the transverse and longitudinal components are equal, one has
\be \Delta^{(1)} \rho = \frac{\sqrt{2} G_F}{ 8 \pi^2} f_1 (s=0, m_a,m_b) \en
\hspace*{-4mm}
\be
f (s,m_a,m_b)= \frac{1}{2} \left[ \Pi_{L}^A (s,m_a,m_a)+\Pi_{L}^A(s,m_b,
m_b) \right] - \Pi_{L}^V(s,m_a,m_b) - \Pi_{L}^A(s,m_a,m_b) \non
\en
Using the previous expressions of $\Pi_{T,L}^{V,A}(0)$ one obtains the
well--known one--loop expression that is quadratic in the fermion masses
\cite{vel}
\be
f_1(0,m_a,m_b) = \frac{3}{2} \left[ m_a^2+ m_b^2+\frac{2m_a^2 m_b^2}{m_a^2
-m_b^2}\log \frac{m_b^2}{m_a^2} \right]
\en

\subsection*{3.~High--momentum transfer}

\renewcommand{\theequation}{3.\arabic{equation}}
\setcounter{equation}{0}

As previously discussed, one can calculate the QCD corrections to the vacuum
polarization functions of the electroweak gauge bosons using a dispersive
approach: the polarization functions can be expressed as dispersive integrals
of their imaginary parts, the expressions of which are available in the
literature at this order of perturbation theory for arbitrary quark masses
\cite{ima}. However, to connect this result with the full expression that can
be derived using dimensional regularization, one needs the high--momentum
transfer behavior of the polarization function in the latter scheme. \s

\nn We have therefore calculated the expressions of the vacuum polarization
functions of the weak gauge bosons at order ${\cal O}(\alpha \alpha_S)$,
Fig.~1b plus the corresponding counterterms, in the limit of high--momentum
transfer [and also at zero--momentum transfer as will be discussed later] for
arbitrary quark masses $m_a \neq m_b \neq 0$ in the on--shell mass scheme. This
scheme, where the quark masses are defined as the poles of their propagators,
is usually used to calculate electroweak radiative corrections \cite{sir}. Here
we will simply give the final result; details on the calculation will be found
in Ref.~\cite{Paolo}.\s

\nn Using the same notation for the vacuum polarization functions as in the
one--loop case, and defining the latter similarly to eq.~(2.3)
\be
\Pi^{ij}_{T,L} (s) \ = \ \frac{\alpha}{\pi} \ \frac{\alpha_S}{\pi} \ \left[ \
v^i v^j \Pi_{T,L}^V(s)+ a^i a^j \Pi_{T,L}^A(s) \ \right]
\en
one obtains for $\Pi_{T,L}^{V,A}(s)$ in the limit $|s| \rightarrow \infty$
\be
\Pi_{T}^{V,A}(|s|\rightarrow \infty) &=& s \ \left[ \frac{1}{2\epsilon} -
\frac{1}{2}(\rho_a+\rho_b)+ \frac{1}{2}(\log \alpha +\log \beta) +\frac{55}{12}
 - 4 \zeta(3) \right] \non \\
&+& (m_a\mp m_b)^2 \left[ \frac{3}{2 \epsilon^2} +\frac{1}{2\epsilon} \left(
\frac{11}{2} -3 \rho_a -3 \rho_b \right) -\frac{11}{4}(\rho_a+\rho_b)
-\frac{11}{8} \right. \non \\
&+& \left. \frac{3}{4}(\rho_a+\rho_b)^2 -\frac{9}{4}(\log\alpha +\log \beta)
-\frac{3}{2} \log \alpha \log \beta +6\zeta(3) \right] \non \\
&+& (m_a^2-m_b^2) \log \frac{m_a^2}{m_b^2} \left[ -\frac{3}{2\epsilon}+
\frac{3}{2} (\rho_a+\rho_b) - \frac{3}{4} (\log\alpha + \log\beta)
+\frac{3}{4} \right] \non \\
&+& 3(m_a^2+m_b^2) (\log \alpha + \log \beta ) \\
\Pi_{L}^{V,A}(|s| \rightarrow \infty)  &=& (m_a\mp m_b)^2 \left[ \frac{3}{2
\epsilon^2} +\frac{1}{2\epsilon} \left(\frac{11}{2} -3 \rho_a -3 \rho_b \right)
-\frac{11}{4}(\rho_a+\rho_b) -\frac{3}{8} \right. \non \\
&+& \left. \frac{3}{4}(\rho_a+\rho_b)^2 -\frac{9}{4}(\log\alpha +\log \beta)
-\frac{3}{2}\log \alpha \log \beta +6\zeta(3) \right] \non \\
&+& (m_a^2-m_b^2) \log \frac{m_a^2}{m_b^2}
\left[ -\frac{3}{2\epsilon}+ \frac{3}{2} (\rho_a+\rho_b) - \frac{3}{4}
(\log\alpha + \log \beta) -5 \right]
\en

\nn with $\alpha= -m_a^2/s ,~\beta =-m_b^2/s$ and $\zeta(3)=1.202$.
At this stage a few remarks are mandatory.

\begin{description}

\item[$i)$] \ \ \ In the previous expressions the momentum transfer is defined
in the space--like region, $s<0$. The continuation to the physical region can
be
straightforwardly obtained by adding a small imaginary part $-i \epsilon$ to
the quark masses squared. This reduces to make the substitutions
$\log(m_{a,b}^2
/-s) \rightarrow \log|m_{a,b}^2/-s| + i\pi$.

\item[$ii)$] \ \ As expected, only the transverse part of the vacuum
polarization function is quadratically divergent for $|q| \rightarrow \infty$.
This divergent term, the expression of which is in agreement with the one
obtained in Ref.~\cite{fnr}, is the same for axial and axial--vector currents
as expected from chiral symmetry. Moreover, as required by the
Kinoshita--Lee--Nauenberg theorem \cite{kln}, this term does not introduce any
mass singularity as $m_{a,b}$ tend to zero: $-(\rho_a+\rho_b)$ and $\log\alpha
+\log \beta$ combine to give $ 2 \log (\mu^2/-s)$.

\item[$iii)$] \ Since the vector part of the longitudinal component should
vanish for $m_a=m_b$, it must be proportional to $(m_a-m_b)$ or
$(m_a^2-m_b^2)$;
and since the axial--vector component can be obtained by changing the sign of
one of the two masses, it must be proportional $(m_a+m_b)$ or $(m_a^2-m_b^2)$
[$\log m_a^2/m_b^2$ alone would have introduced mass singularities]. This
behavior is explicitely exhibited by the previous expression of $\Pi_{L
}^{V,A}(s)$.
\end{description}

\vspace*{0.5cm}

\nn Let us now write the real parts of $\Pi_{T}^{V,A}(s)$, analogously to
eq.~(2.10),
\be
\Pi_{T}^{V,A}(s)= s \tilde{X}_{2}+(m_a \mp m_b)^2 \tilde{Y}_{2}+(m_a^2-m_b^2)
\log{m_a^2\over m_b^2} \ \tilde{Z}_{2}+ F_2^{V,A} (s)
\label{dimreg}
\en
where the two--loop divergent constants $\tilde{X}_2,~\tilde{Y}_2$ and $\tilde{
Z}_2$ involve only the poles in $\epsilon$ and the logarithms of the scale
$\mu$ of the expressions of eq.~(3.3); they are given by
\be
\tilde{X}_2 &=& {1 \over 2\epsilon} - \log\frac{m_am_b}{\mu^2} \non \\
\tilde{Y}_2 &=& {3 \over 2\epsilon^2} + {1\over\epsilon} \left({11\over4} -3
\log\frac{m_am_b}{\mu^2}  \right) -\frac{11}{2}\log\frac{m_am_b}{\mu^2}
+3\log^2\frac{m_am_b}{\mu^2} \non \\
\tilde{Z}_2 &=& - {3 \over 2\epsilon} + 3\log\frac{m_am_b}{\mu^2}
\en

\nn Using the results of the imaginary parts of the vector bosons vacuum
polarization functions at ${\cal O}(\alpha \alpha_S)$ given in Ref.~\cite{ima}
for arbitrary quark masses, one can express the result of the dispersive
integral eq.~(2.7) in this case as
\be
\Pi_{T}^{V,A} (s) &=& sX_2 +(m_a \mp m_b)^2 Y_2 + (m_a^2-m_b^2) \log{m_a^2
\over m_b^2} \ Z_2 + F_2^{V,A}(s) \non \\
& & + \ \frac{1}{2\pi i} \oint_{|s'|=\Lambda^2}\frac{{\rm d}s'}{s'} \left[
\Pi_{L}^{V,A} (s') -s\Phi^{V,A} (s') \right] \label{disp}
\en

\nn where the finite function\footnote{Note that the definition of $F_{1,2}^{V,
A}(s)$ used here differs from the one used in Ref.~\cite{ks}, since we
normalize the divergent constants to the dimensional regularization result
eq.~(3.2) keeping only the ultraviolet divergencies and the related
logarithms.}
$F^{V,A}_2 (s)$ should be the same as the one in eq.~(3.4). This has been
verified up to ${\cal O}(\alpha \alpha_S)$ \cite{kn} in the special cases for
which both approaches were used, i.e. $m_a=m_b$ and $m_b=0$. \s

\nn Here again, since the result of eq.~(3.6) should be consistent with the one
obtained in dimensional regularization, the integral over the large circle must
be absorbable in a redefinition of the divergent constants $X_2,Y_2$ and $Z_2$.
Indeed, when we use dimensionally regularized expressions for $\Phi(s)$ and
$\Pi_L(s)$, the integral involving $\Phi(s)$ in eq.~(3.4) will replace $X_2$ by
$\tilde{X_2}$, and the one involving $\Pi_L(s)$ will replace $Y_2$ and $Z_2$ by
respectively $\tilde{Y}_2$ and $\tilde{Z}_2$. One would have
\be
\frac{1}{2\pi i}\oint_{|s'|=\Lambda^2}ds' \frac{\Phi^{V,A}(s')}{s'} &=&
X_2-\tilde{X}_{2} \non \\
\frac{1}{2\pi i} \oint_{|s'|=\Lambda^2}ds' \frac{\Pi^{V,A}_{L}(s')}{s'}
&=& (m_a \mp m_b)^2 ( \tilde{Y}_{2} -Y_2) \non \\
&+& (m_a^2-m_b^2) \log{m_a^2\over m_b^2} \ (\tilde{Z}_{2}-Z_2)
\en

\nn Using the ${\cal O}(\alpha \alpha_S)$ results of the imaginary parts of the
vacuum polarization functions given in \cite{ima}, and after a straightforward
computation, we obtain for the two--loop constants $X_2,Y_2$ and $Z_2$
\be
X_2 &=& \log\frac{\Lambda^2}{m_am_b} + 4\zeta(3) - {55\over12} \nonumber\\
Y_2&=&{3\over2} \log^2 \frac{\Lambda^2}{m_am_b} -{9\over2} \log\frac{\Lambda^2}
{m_am_b}-{3\over2}\log^2{m_a\over m_b}+{3\over8}+6\zeta(3)-{\pi^2\over2} \non
\\
Z_2&=&-{3\over2} \log\frac{\Lambda^2}{m_am_b} +5
\en
The divergent part [as $\Lambda$ tends to infinity] of these constants for
arbitrary masses, as well as their expressions in the special cases $m_a=m_b$
and $m_b=0$ have been derived in Ref.~\cite{kn}. \s

\nn As a consequence of the use of eq.~(3.7) in the dispersion relation
eq.~(2.7), one can now implement in the general case of arbitrary quark masses,
non--perturbative effects \cite{ks,ksf} in the absorptive part of the amplitude
[that can be calculated by integrating, e.g. numerically, the expressions of
the imaginary parts of the polarization functions derived in Ref.~\cite{ima}]
defined not only in the on--shell scheme, which is natural in the dispersive
approach, but also in the ${\rm \overline{MS}}$ scheme for which a dimensional
regularization calculation is required.

\subsection*{4.~Zero--momentum transfer}

\renewcommand{\theequation}{4.\arabic{equation}}
\setcounter{equation}{0}

We come now to the discussion of the zero--momentum transfer behavior of the
vacuum polarization functions. In this limit, the evaluation of the two--loop
diagrams Fig.~1b in the on--shell mass scheme, leads to the following
expression for $ \Pi_{T,L}^{V,A}(0)$
\be
\Pi_{T,L}^{V,A} (0) & = & \frac{3}{2\epsilon^2} (m_a^2+m_b^2)+
\frac{1} {\epsilon} \left[ \frac{11}{4}(m_a^2+m_b^2)-3m_a^2 \rho_a-3m_b^2
\rho_b \right]  - \frac{11}{2} (m_a^2 \rho_a+ m_b^2\rho_b) \non \\
&+ &  3m_a^2 \rho_a^2 +3m_b^2 \rho_b^2 + \frac{35}{8}(m_a^2+m_b^2) +\frac{1}{4}
(m_a^2-m_b^2) \left [G\left(\frac{m_a^2}{m_b^2} \right) -G\left( \frac{m_b^2}
{m_a^2} \right) \right] \non \\
&+& \frac{m_a^2m_b^2}{m_a^2-m_b^2} \log \frac{m_a^2}{m_b^2}
+ m_a^2 m_b^2 \frac{m_a^2+m_b^2}{(m_a^2-m_b^2)^2} \log^2
\frac{m_a^2}{m_b^2} \non \\
&\pm  & \ m_a m_b \ \left[ -\frac{3}{\epsilon^2}+ \frac{1}{\epsilon} \left(
3\rho_a + 3\rho_b-\frac{11}{2} \right) -\frac{3}{2} (\rho_a + \rho_b)^2 +
4 (\rho_a + \rho_b) -\frac{31}{4} \right. \non \\
&+ &  \left. 3 \frac{m_a^2 \rho_b- m_b^2\rho_a} {m_a^2-m_b^2} +
3\frac{m_a^2m_b^2} {(m_a^2-m_b^2)^2} \log^2 \frac {m_a^2}{m_b^2} \ \right]
\en

\nn where, in terms of the Spence function defined by ${\rm Li}_2(x)= -\int_0^1
y^{-1} \log (1-xy) {\rm d}y$, the function $G$ is given by
\be
G(x) = 2{\rm Li}_2(x) +2 \log x \log (1-x)+ \frac{x}{1-x} \log^2 x
\en

\nn As one might have expected, there is no singularity in the self--energies
in this limit. In addition, besides the manifest symmetry in the exchange $m_a
\leftrightarrow m_b$, the previous expressions exhibit the facts that
$\Pi^{V,A}_{T,L}(0)$ can be obtained from $\Pi^{A,V}_{T,L}(0)$ by simply making
the substitution $m_a(m_b) \rightarrow -m_a(-m_b)$ as expected from $\gamma_5$
reflection symmetry and that the longitudinal and transverse components are
equal. These features provide good checks of the calculation. \s

\nn Using the previous expression, one readily obtains the QCD corrections to
the contribution of a heavy quark isodoublet to the $\rho$ parameter in the
general case $m_a \neq m_b \neq 0$. Defining the contribution to the $\rho$
parameter analogously to eq.~(2.15)
\be
\Delta^{(2)} \rho = \frac{\sqrt{2} G_F}{8 \pi^2} \frac{\alpha_s}{\pi} \ f_2
(0,m_a,m_b)
\en
the function $f_2$ will be given by
\be
f_2(0,m_a,m_b) &=& - 3 \left\{ m_a^2+m_b^2  +\ 2\ \frac{m_a^2 m_b^2}
{m_a^2-m_b^2}\log\frac{m_a^2} {m_b^2} \left[ 1+ \frac{m_a^2+m_b^2}{m_a^2-m_b^2}
\log\frac{m_b^2}{m_a^2} \right] + (m_a^2-m_b^2) \right. \non \\
&& \times \left. \left[ 2{\rm Li_{2}} \left( \frac{m_a^2}{m_b^2} \right)+2 \log
\frac{m_a^2}{m_b^2} \log \left( 1-\frac{m_a^2}{m_b^2} \right) - \frac{m_a^2}{
m_a^2-m_b^2} \log^2 \frac{m_a^2}{m_b^2} - \frac{\pi^2}{3} \right] \right\}
\hspace*{3mm}
\en

\nn $f_2(0,m_a,m_b)$ is free of ultraviolet divergences as it should be since
$\Delta \rho$ is an observable physical quantity; furthermore it does not
depend on the 't Hooft mass scale $\mu$. Note that the symmetry in the
interchange of $m_a$ and $m_b$ is now hidden in the term in the last line
of the previous equation, but this term is simply $G(m_a^2/m_b^2)-G(m_b^2/
m_a^2)$ and we have used the fact that $G(1/x)+G(x)=2\pi^2/3$. \s

\nn In the limit of large mass splitting between the two quarks, $m_a \gg m_b$,
the QCD corrections to the $\rho$ parameter reduce to the known result
for $m_b=0$ \cite{moi}
\be
f_2(0,m_a,0) = - \frac{2}{3} \ f_1(0,m_a,0) \ \frac {\alpha_S}{\pi}
\left( 1+ \frac{\pi^2}{3} \right)
\en

\nn The contribution of a new quark isodoublet to the $\rho$ parameter is
exemplified in Fig.~2a both in the one--loop approximation [full line] and
including the QCD radiative corrections [dashed line]. We have fixed the mass
of one quark to 250 GeV and varied the mass of the other quark from 100 to 400
GeV; for the strong coupling constant we have used the numerical value
$\alpha_S \simeq 0.12$. [The scale at which $\alpha_S$ should be evaluated is
the mass of the heavy quark \cite{SVZ}; this choice can also be justified by
arguments based on effective field theory \cite{GR}.] \s

\nn As one can see the contribution of the heavy quark isodoublet to $\Delta
\rho$ is very small for approximately mass degenerate quarks, $m_a=m_b$; this
illustrates the well known fact that large contributions to $\Delta \rho$ are
induced only if the splitting between the two quark masses is large. We see
that the QCD corrections are always negative and therefore reduce the size of
$\Delta \rho$. In fact, in this mass range the QCD corrections are practically
constant and amount to a decrease of $\Delta \rho$ by approximately 15\% as
can be seen from Fig.~2b. \s

\nn To close this section, let us briefly discuss the application of the
dispersive method to physical observables, taking the example of the fermionic
contribution to the $\rho$ parameter. Using the fact that at zero--momentum
transfer, the transverse and longitudinal components of the $W$ and $Z$
self--energies are equal, and recalling that the vector component vanishes in
this limit for equal mass quarks, we have written $\Delta \rho$ in terms of
the function $f(s,m_a,m_b)$ defined in eq.~(2.5). Since $\Delta \rho$ is an
observable physical quantity [and therefore free of ultraviolet divergences],
and since the divergences of $\Pi_L$ are independent of $s$, $f(s,m_a,m_b)$ can
be evaluated by means of an unsubtracted dispersion relation
\be
f(s,m_a,m_b)=  {1\over\pi} \int_{(m_1+m_2)^2}^{\Lambda^2} ds' \frac{{\cal I}m
f(s',m_a,m_b)}{s' -s-i\epsilon}
\en
We have verified at the two--loop level, that the ultraviolet finite function
$f(s, m_a, m_b)$ tends to zero as $s$ becomes infinite and therefore satisfies
the condition
\be
\lim_{\Lambda^2\to \infty} \ \oint_{|s'|=\Lambda^2} \ {ds'\over s'} \
f(s', m_a, m_b) \ = \  0
\en
\nn which generalizes the result obtained in Ref.~\cite{ks} in the case of
one massless quark, $m_b=0$.

\subsection*{5. Summary}

In this paper, we have derived the asymptotic behavior of the heavy quark
contribution to the electroweak gauge bosons vacuum polarization functions at
${\cal O}(\alpha \alpha_S)$ for arbitrary quark masses, $m_a \neq m_b \neq 0$.
\s

\nn Using the high--momentum transfer behaviour, we have extended to the case
of arbitrary fermion masses at the two--loop level, recent analyses of the
connection between calculations using dimensional regularization and the ones
performed by means of dispersive integration. One can therefore implement, in
the general case, non--perturbative effects in the absorptive part of the
vacuum polarization function [which can be calculated using the available
expressions of the imaginary parts] not only in the on--shell scheme but also
in the ${\rm \overline{MS}}$ scheme which inherently requires a dimensional
regularization calculation. \s

\nn With the results that we obtain for the zero--momentum transfer behavior of
the electroweak gauge bosons self--energies, we derived the QCD corrections to
the leading contribution of new heavy quarks to electroweak observables: the
deviation of the $\rho$ parameter from unity. This correction is practically
constant and amounts to a decrease of the one--loop value of $\Delta \rho$ by
approximately 13--15\% for $\alpha_S \sim 0.12$.

\vspace*{1cm}

\nn {\bf Acknowledgements.} We thank Professor Alberto Sirlin for suggesting
this work and for useful discussions. Technical help from I.~Cherchneff, C.
Parrinello and J. Papavassilou is gratefully acknowledged.
This work is partially supported by the National Sciences and Engineering
Research Council of Canada.

\vspace*{2cm}

\subsection*{Figure Captions}

\vspace*{.5cm}

\begin{description}

\item[Fig.1] Contribution of a quark pair to the vacuum polarization function
of a gauge boson in the one--loop approximation (1a) and at the two--loop level
(1b). \s

\item[Fig.2] Contribution of a heavy quark isodoublet to the $\rho$ parameter.
(2a) At ${\cal O}(\alpha)$ [full line] and total contribution at ${\cal O}
(\alpha \alpha_S)$ [dashed line], (2b) total contribution at ${\cal O} (\alpha
\alpha_S)$ normalized to the one at ${\cal O}(\alpha)$. We have taken $\alpha_S
\simeq 0.12$.

\end{description}

\end{document}